\documentclass{sig-alternate-05-2015}
\usepackage{color}

\begin{document}

\setcopyright{acmcopyright}

\newtheorem{definition}{Definition}
\newcommand{\col}[1]{\textsf{#1}}

\doi{10.475/123_4}

\isbn{123-4567-24-567/08/06}



%

\title{Data Compression for Analytics over Large-scale In-memory Column Databases}

\author{\vspace{0.3cm}Chunbin Lin~~~~Jianguo Wang~~~~Yannis Papakonstantinou\\
\affaddr{Department of Computer Science and Engineering}\\
\affaddr{University of California, San Diego}\\
\affaddr{\{chunbinlin, csjgwang, yannis\}@cs.ucsd.edu}\\
}

\maketitle
\begin{abstract}
Data compression schemes have exhibited their importance in column databases by contributing to the  high-performance OLAP (Online Analytical Processing) query processing. Existing works mainly concentrate on evaluating compression schemes for disk-resident databases as data is mostly stored on disks. With the continuously decreasing of the price/capacity ratio of main memory, it is the tendencies of the times to reside data in main memory.
But the discussion of data compression on in-memory databases is very vague in the literature.
In this work, we present an updated discussion about whether it is valuable to use data compression techniques in memory databases.
If yes, how should memory databases apply data compression schemes to maximize performance?

\end{abstract}

\section{Introduction}
Data compression is a well-known optimization in traditional disk-resident database systems~\cite{abadi2006integrating,chen2001query,IdreosGNMMK12}.
The most important benefit of compression is to reduce the expensive disk I/O cost significantly by reducing the data size.
The CPU decompression overhead (if needed) is inarguably negligible compared to the I/O time saving because of the giant performance gap between disks and CPUs.
As a result, modern databases systems all embrace compressions, e.g., Microsoft SQL Server, IMB DB2, Oracle, PostgreSQL.




However, recent years have witnessed an explosion of main memory databases, e.g., MonetDB~\cite{IdreosGNMMK12} and SAP HANA~\cite{FarberMLGMRD12}. It is fueled by several fundamental technology trends.
(1) Memory is $5\sim6$ orders of magnitude faster than disk-based storage, which is crucial to high-performance big data analytics.
(2) The capacity of memory is continuously increasing while the price (in \$/GB) is dropping. As a result, today's commodity server machines have more than 32 GB's memory.
(3) The breakthrough in non-volatile main memory exemplified by STT-RAM, PCM, and Re-RAM \cite{zhang2015} makes it possible to extend the DRAM capacity cost-effectively. That is because NVMM is cheaper than DRAM (in \$/GB) while still being competitive in terms of performance.
(4) The significant advances in networking also makes it feasible to extend DRAM capacity by connecting many machines together through ultra-fast networks, e.g., RAMCloud \cite{OusterhoutGGKLM15} is such a memory system.

With everything in memory and disks being removed from the critical path, the decompression overhead pumps out, which adversely affects the performance of in-memory databases. Thus, a natural question is that, \textbf{\emph{is it still reasonable to apply data compression for in-memory databases}?} If the answer is yes, then the follow-up question is \textbf{\textit{how to apply data compression schemes for in-memory databases in order to maximize the performance?}}

Existing works provide  insufficient discussion on the above two questions. Therefore, in this work, we want to bridge the gap. In particular, we want to answer the two questions with a focus on OLAP (Online Analytical Processing) workloads\footnote{The counterpart is OLTP (Online Transaction Processing) workloads.} due to their importance in decision support and data mining applications. OLAP workloads  usually require reading a large portion of data from particular columns. Therefore, it is preferred to use column-oriented databases to answer such queries efficiently.\footnote{It has been proven that column-oriented databases outperform row-oriented databases on OLAP workloads by $1\sim2$ orders of magnitude~\cite{abadi2006integrating}.}  Therefore, this work mainly discusses data compression for OLAP workloads over large-scale in-memory column databases.

\section{To compress or not in memory databases?}
In this section, we elaborate whether it still makes sense to use data compression in memory-based databases.

An obvious advantage of data compression in memory databases is less memory requirement. Next, we explain more about system performance.

Intuitively, if the entire data is stored in memory, then data compression will slow down the performance due to the additional decompression overhead.
This is indeed true for some compression schemes, e.g., Huffman encoding~\cite{Huffman52}.
However, for some other compression schemes (e.g., dictionary encoding~\cite{chen2001query}, run-length encoding~\cite{abadi2006integrating} and bitmap encoding~\cite{wu2006optimizing}), as we will show in Section~\ref{sec:compression}, they allow queries to be answered directly over compressed data without decompression at all.
More importantly, performing queries over compressed data directly is even faster than that over uncompressed raw data, since less data is processed.

Therefore, whether a data compression scheme is useful in memory databases depends on whether it can answer queries directly without decompression. If yes, it is strongly recommended to use that compression in memory databases, because it ``kills two birds with one stone'', saving not just memory requirement, but also improving query performance. Otherwise, it is not recommended (since the decompression time can be a dominant factor if data is huge), although it is up to the applications to balance the tradeoff between the decompression overhead and the saving of memory requirement.

\section{How to compress in memory \\databases?}
\label{sec:compression}

In this section, we (i) introduce the widely used database compression schemes, (ii) analyze whether they can be applied in memory databases, and (iii) propose our optimizations (if any) to further improve the performance.



\subsection{Dictionary encoding (DICT)}
\noindent\textbf{Compression description.}
Dictionary encoding (DICT)~\cite{chen2001query} is widely used for string-type columns. It maps each original string value to a 32-bit integer (a.k.a \textbf{word-DICT}) according to a global dictionary table.  For example, in a \col{State} column,  ``Alabama'' is mapped to ``1'' and ``Alaska'' is mapped to ``2''. Note that, one underlying assumption is that the global dictionary table is small in order to be fitted in memory. 



\begin{figure}[h!]
\centering
\includegraphics[width=1\columnwidth]{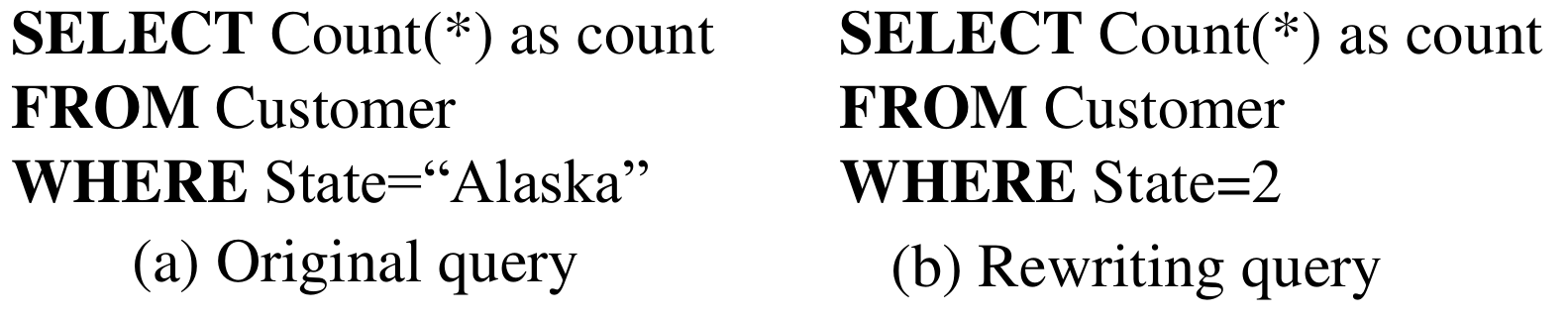}
\caption{\textbf{Example query and its rewriting query}}\label{fig:dictionary_example}
\end{figure}

\noindent\textbf{Is it useful in memory databases?}
Dictionary encoding schemes can be applied in memory databases, since they can answer queries directly on the compressed data by encoding query conditions using the same global dictionary table. For example, assume the \col{State} column of a \col{Customer} table is encoded by word-DICT, then SQL query of ``counting the total number of customers in Alaska'' (shown in Figure~\ref{fig:dictionary_example}(a)) is rewritten to the query in Figure~\ref{fig:dictionary_example}(b) by converting the query condition ``Alaska'' to ``2''. 


The rewriting query can be performed directly over the compressed column, which is much faster than that over the uncompressed data, because it reduces the expensive string comparison problem to the cheap integer comparison problem.

\noindent\textbf{Our optimizations.}
Existing dictionary encoding schemes mainly focuses on string-type columns, however, we want to emphasize that, there are some optimization opportunities for integer-type columns, which is common for foreign keys.\footnote{Usually, if a foreign key column is string-type, then a word-aligned dictionary encoding can be applied to convert it to an integer-type column.}

Integer values can be mapped to smaller codes with the minimal number of bits to distinguish the original integers. More formally, let $D$ be the domain size, then each original value can be encoded in $\lceil \log_{2} D\rceil$ bits, which is called \textbf{bit-DICT}. For example, suppose the domain size $D=50$, each value can be expressed by $\lceil \log_{2} 50\rceil = 6$ bits, then number 8 and 22 can be encoded as 001000 and 010110 respectively.\footnote{The most significant 0's are padding 0's.} 

It is worth mentioning that the global dictionary table is no longer needed because the mapping from integer to its binary expression (with padding 0's) is a one-to-one mapping. Note that, maintaining the global table is a limitation of existing dictionary encoding schemes when applying to string-type columns, because the dictionary table is big when the domain size is large.


\noindent\textbf{Remark.} For in-memory column databases, word-DICT is a reasonable choice for string-type columns with small domain size (say less than $50,000$~\cite{abadi2006integrating}). And, our bit-DICT is highly recommended for integer-type columns (e.g., foreign key columns).


\subsection{Run-length encoding (RLE)}
\noindent\textbf{Compression description.} Run-length encoding (RLE)~\cite{abadi2006integrating} is an attractive approach for compressing data in column databases. It compresses continuous duplicated values to a compact singular representation, which implies that it is only applicable to sorted columns. Traditional RLE expresses the repeats of the same value as pairs (\textbf{v}alue, run-\textbf{l}ength), we call it \textbf{vl-RLE}. For example, value ``Alaska'' appears 10000 times continuously in the \col{State} column, then it can be simply expressed as (Alaska, 10000) instead of storing 10000 duplicates.

\noindent\textbf{Is it useful in memory databases?} The answer is absolutely positive.  vl-RLE is efficient for queries that operate on the compressed column. For instance, the query in Figure~\ref{fig:dictionary_example}(a) can be answered directly on the compressed column, i.e., the answer is 10000 in this case. Not only the vl-RLE can be applied in memory databases, but also its two variants: \textbf{vsl-RLE} and \textbf{vs-RLE}.

vsl-RLE uses a (\textbf{v}alue, \textbf{s}tart-position, run-\textbf{l}ength) scheme to represent the repeats of the same value. By adding the \textit{start-position}, it tells the start row id of the repeat values. The queries that can be directly executed on the compressed data by vsl-RLE is a superset of that by vl-RLE, since it supports queries accessing other columns. Consider a table R(A, B) with A encoded by vsl-RLE, the encoding for values $t\in A$ is $(t, s, l)$, where $s$ is the start row id and $l$ is the number of repeats of $t$. The answer of query $\pi_B\sigma_{A=t}R$ is a list of values in column B from $s$-th row to $(s+l)$-th row.

vs-RLE uses a (\textbf{v}alue, \textbf{s}tart-position) scheme to represent the repeats of the same value. It has the same capability with vsl-RLE of answering queries directly over compressed data, but needs less memory requirement. Suppose the table R(A, B) with A encoded by vs-RLE, the encoding for values $t\in A$ and $t+1\in A$ are $(t, s)$ and $(t+1, s')$ respectively. The answer of query $\pi_B\sigma_{A=t}R$ is a list of values in column B from $s$-th row to $(s'-s)$-th row. 

%
%
%
%

\noindent\textbf{Our optimizations.}
We claim that for integer-type columns, vs-RLE can be further compressed by bit-DICT, we call it \textbf{vsb-RLE}. Due to the consideration of the query processing performance, only the values (not run-lengths) are further encoded by bit-DICT. Note that, as we described before, bit-DICT does not hurt performance, thus vsb-RLE has similar performance with vs-RLE and vsl-RLE but requires less space. Table~\ref{table:RLE} reports the space cost of each RLE.

\begin{table}[h!]
\small
\centering
\renewcommand{\tabcolsep}{2.8mm}
       \begin{tabular}{c|c|c|c|c}\hline\hline
       Uncompress & vl-RLE & vsl-RLE & vs-RLE & vsb-RLE \\\hline
              4GB & 8MB    & 12MB    & 8MB    & \textbf{6.5MB}\\\hline \hline
        \end{tabular}
\caption{\textbf{Space cost of different RLE schemes on an integer-type column with 1 billion rows and 1 million distinct values}}
 \label{table:RLE}
\end{table}

\noindent\textbf{Remark.} For sorted columns, vsb-RLE is highly recommended, because it has less space than other competitors but with similar performance.

\subsection{Bitmap encoding}
\noindent\textbf{Compression description.}
Another well-studied encoding method in column databases is bitmap encoding~\cite{antoshenkov1995byte,wu2006optimizing}. Each distinct value $t$ is associated with a bit-vector indicating the occurrences of $t$ on the column. The default values in the bit-vector are zeros while the $i$-th position of the bit-vector is set to 1 if the $i$-th position has value $t$ on the original column.

The size of each bit-vector is the size of the table, which is extremely large in large-scale databases. To reduce the space overhead of the bitmap index, run-length encoding is employed to compress the continuous 1's and 0's on bit-vectors. Two representative compressed bitmap encodings are BBC (byte-aligned bitmap code)~\cite{antoshenkov1995byte} and WAH (word-aligned hybrid code)~\cite{wu2006optimizing}. Compared with BBC, WAH outperforms BBC by about 12 times and uses about 60\% more space~\cite{wu2006optimizing}.

Though both WAH and BBC can reduce the size of each bit-vector, the number of bit-vectors cannot be reduced. Therefore, compressed bitmap schemes are only applicable to columns with small size, e.g., less than 50~\cite{abadi2006integrating}.

\noindent\textbf{Is it useful in memory databases?}
Bitmap can be applied in memory databases, since it allows logical bitwise operation directly on a compressed bitmap~\cite{wu2006optimizing}. For example, consider the query:

\smallskip
~~~~~~~\textbf{SELECT} B \textbf{FROM}  R \textbf{WHERE}  A $\leq$ 2
\smallskip

The main operation of the query is to get row ids satisfying $\sigma_{A\leq2}R$, which can be answered by performing a bitwise logical operation $b_1~OR~b_2$, where $b_1$ and $b_2$ are compressed bit-vectors for value 1 and 2 respectively.

\noindent\textbf{Remark.} It is recommended to use compressed bitmap for a column in memory column databases only when the domain size is small.

\subsection{Huffman encoding}
\noindent\textbf{Compression description.}
Huffman encoding~\cite{Huffman52} is a representative of the variable length encodings, which is widely used in many areas. It is based on the frequency of occurrence of a data. The principle is to use a smaller number of bits to encode the data that occurs more frequently.

\noindent\textbf{Is it useful in memory databases?}
Unfortunately, Huffman encoding cannot be applied in memory databases, since it needs to decompress the whole column to answer queries, which is time-consuming. This is because Huffman encoding does not support partial decompression due to its variable length structure. We conducted experiments to test the decompression performance. The decompression time is $3.2$ minutes for a column with 1 million rows (domain size is 100 and data follows Zipf distribution). 


%

\noindent\textbf{Remark.} We do not recommend to use Huffman encoding  for in-memory column databases.

\vspace{1mm}
\section{Conclusion}

\begin{figure}[htbp]
  \centering
  \includegraphics[width=1\columnwidth]{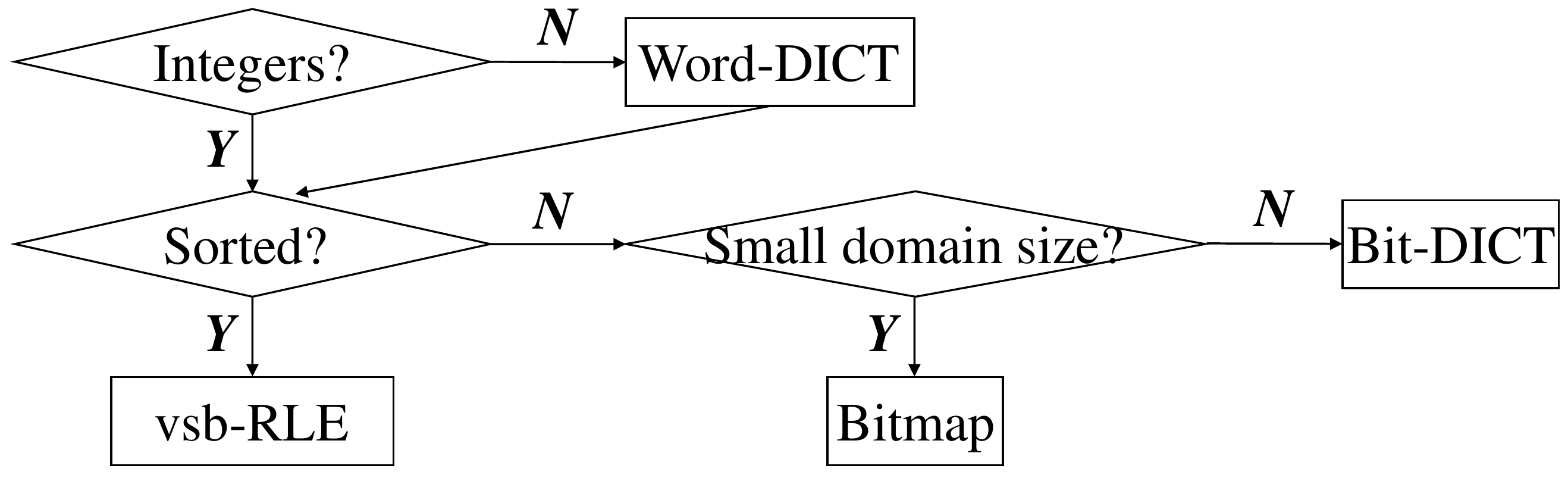}
\caption{A decision tree of selecting proper compression schemes}\label{fig:decision}
\end{figure}

In this paper, we present an updated discussion of commonly used data compression schemes in memory column databases. This study contributes the following messages to our community:
\begin{itemize}
	\item It is still beneficial to apply data compression for in-memory databases. A clear benefit is that, compression can save memory requirement. More importantly, some compressions (e.g., dictionary encoding, run-length encoding and bitmap) can directly answer queries over compressed data, yielding high performance.
	\item We give insights regarding how to use data compression in memory databases to maximize performance, as shown in  Figure~\ref{fig:decision}. If a column is integer-type (if not, it is recommended to apply word-aligned dictionary encoding first to transform to integer-type), then if the column is sorted, we recommend to use vsb-RLE. Otherwise, depending on the domain size, we can use compressed bitmap encoding (small domain size) or bit-aligned dictionary encoding (big domain size).

    \item Huffman encoding is not recommended to be used in memory databases due to the slow decompression performance.
\end{itemize}

{\small
\bibliographystyle{abbrv}
\bibliography{ref}  
}
\end{document}